\documentclass[showpacs,10pt,twocolumn,prl]{revtex4-1}

\usepackage{amsmath}
\usepackage{amssymb}
\usepackage{graphicx}
\usepackage{amssymb}
\usepackage{graphics}
\usepackage{epsfig}
\usepackage{CJK}
\usepackage{color}

\begin{document}

\begin{CJK*}{GBK}{}
\title{Electronic transport and thermoelectricity in selenospinel Cu$_{6-x}$Fe$_{4+x}$Sn$_{12}$Se$_{32}$}
\author{Yu Liu,$^{1,*}$ Zhixiang Hu,$^{1,2}$ Xiao Tong,$^{3}$ David Graf,$^{4}$ and C. Petrovic$^{1,2}$}
\affiliation{$^{1}$Condensed Matter Physics and Materials Science Department, Brookhaven National Laboratory, Upton, New York 11973, USA\\
$^{2}$Materials Science and Chemical Engineering Department, Stony Brook University, Stony Brook, New York 11790, USA\\
$^{3}$Center of Functional Nanomaterials, Brookhaven National Laboratory, Upton, New York 11973, USA\\
$^{4}$National High Magnetic Field Laboratory, Florida State University, Tallahassee, Florida 32306-4005, USA}

\date{\today}

\begin{abstract}
We report a study of selenospinel Cu$_{6-x}$Fe$_{4+x}$Sn$_{12}$Se$_{32}$ ($x$ = 0, 1, 2) single crystals, which crystalize in a cubic
structure with the $Fd\overline{3}m$ space group, and show typical semiconducting behavior. The large discrepancy between the activation energy for electrical conductivity $E_\rho$ (32.3 $\sim$ 69.8 meV), and for thermopower $E_\textrm{S}$ (3.2 $\sim$ 11.5 meV), indicates a polaronic transport mechanism between 350 and 50 K. With decreasing temperature, it evolves into variable-range hopping conduction. Furthermore, the heat capacity shows a hump around 25(5) K and diverges from the Debye $T^3$ law at low temperatures, indicating the observation of structural glassy features in these crystalline solids.
\end{abstract}

\maketitle
\end{CJK*}

\section{INTRODUCTION}

New transition metal chalcogenides may enable transformative changes in thermoelectric energy creation and conversion
\cite{RoychowdhuryS,OuyangY,BanikA}. The Kondo-insulator-like semiconductor FeSb$_2$ hosts not only strong electronic correlation but also the highest thermoelectric power factor in nature along with thermopower up to $\sim$ 45 mV K$^{-1}$
\cite{Petrovic1,Petrovic2,HuR2,Bentien,Jie,Takahashi}. The ternary CoSbS also features high value of thermopower $\sim$ 2.5 mV K$^{-1}$
around 40 K but also relatively large value of thermal conductivity $\sim$ 100 W K$^{-1}$ m$^{-1}$ near the thermopower peak temperature
\cite{DuQ}. This inhibits the thermoelectric figure of merit $ZT = (S^{2}$/$\rho$$\kappa$)$T$, where $S$ is thermopower, $\rho$ is
electrical resistivity, $\kappa$ is thermal conductivity, and $T$ is temperature, respectively. All these parameters are interdependent
and related to the electronic structure and charge carrier scattering of the material. To improve a material for thermoelectricity it is important to achieve high thermoelectric power factor ($S^2/\rho$) and low thermal conductivity $\kappa$.

Quaternary materials, especially transition metal chalcogenides, are more tunable than binary and ternary compounds \cite{HanC}.
Quaternary spinel Cu$_6$Fe$_4$Sn$_{12}$Se$_{32}$ features a large slope of density of states near the Fermi energy, favorable for high
value of thermopower \cite{Suekuni3}. This compound crystallizes in cubic yet complex crystal structure of high symmetry with many atoms where high degeneracy of band extrema are commonly found \cite{DiSalvoF,LarsonP,SootsmanJ,SnyderG}. Numerous atoms in the high symmetric unit cell along with the Fe/Sn site disorder contribute to relatively low thermal conductivity $\kappa$ in Cu$_y$Fe$_4$Sn$_{12}$Se$_{32}$
polycrystals \cite{Suekuni4,Suekuni5}. In addition, Cu$_y$Fe$_4$Sn$_{12}$Se$_{32}$ is stable at $y$ = 6 without impurity, where a large $S$ of $\sim$ 200 $\mu$V K$^{-1}$ and a rather low $\rho$ coexist at 340 K as well as low $\kappa$ of $\sim$ 1.5 W K$^{-1}$ m$^{-1}$ at the room temperature.

Here we fabricated a series of Cu$_{6-x}$Fe$_{4+x}$Sn$_{12}$Se$_{32}$ ($x$ = 0, 1, 2) single crystals to shed light on the electronic transport mechanism. X-ray photoemission spectroscopy (XPS) measurements were performed to investigate the valencies of interpolated cations. We observe a relatively high value of thermopower $\sim$ 0.18 mV/K at 300 K for $x$ = 0, which increases to $\sim$ 0.36 mV/K for $x$ = 2. The large discrepancy between the activation energy for electrical conductivity and for thermopower above 50 K points to the polaronic transport mechanism. With decreasing temperature, polaronic transport evolves into the dominant variable-range hopping (VRH) mechanism.

\section{EXPERIMENTAL DETAILS}

Single crystals of Cu$_{6-x}$Fe$_{4+x}$Sn$_{12}$Se$_{32}$ ($x$ = 0, 1, 2) were grown by melting stoichiometric mixture of Cu and Fe (4N, Alfa Aesar) powder, Sn and Se (5N, Alfa Aesar) shot \cite{Suekuni3}. The starting materials were mixed and loaded in 1 cm diameter Al$_2$O$_3$ crucibles, vacuum-sealed in quartz tubes, heated to 650 $^\circ$C over 10 h and dwelled for 12 h, then cooled to 550 $^\circ$C at a rate of 1 $^\circ$C/h, and finally quenched in ice water. Crucible-size-limited single crystals were obtained with shining surface and confirmed by back-reflection Laue x-ray photographs.

The powder x-ray diffraction (XRD) data were taken on crushed crystals with Cu $K_{\alpha}$ ($\lambda=0.15418$ nm) radiation of a Rigaku
Miniflex powder diffractometer. The element analysis was performed using energy-dispersive x-ray spectroscopy (EDX) in a JEOL LSM-6500
scanning electron microscope. Multiple points on several samples were examined and the average actual chemical composition of samples are listed in Table 1. XPS measurements were carried out in an ultrahigh-vacuum (UHV) system with base pressure of $\thicksim2\times$10$^{-10}$ Torr and equipped with a hemispherical electron energy analyzer (SPECS, Phoibos 100) and twin anode x-ray source (SPECS, XR50). Al-K$_{\alpha}$ (1486.6 eV) radiation was used at 15 kV and 20 mA. The angle between the analyzer and x-ray source is 45$^\circ$ and photoelectrons were collected along the sample surface normal. In order to remove potential surface contaminations and oxygen layers, each samples were cleaned in UHV by Ar$^+$ sputtering for 60 minutes under conditions of Ar gas pressure of $2\times$10$^{-5}$ Torr and Ar$^+$ kinetic energy of 2 keV. XPS data was analyzed using Casa XPS and peak positions were calibrated using residual adventitious carbon C 1s at 284.8 eV. The electrical resistivity, thermopower, and thermal conductivity were measured in the quantum design thermal transport option (TTO) in PPMS-9 with standard four-probe method and in continuous measuring mode. The maximum heater power and period were set as 50 mW and 1430 s along with the maximum temperature rise of 3$\%$. The crystals were cut and polished into rectangular bars with typical dimensions of $4\times1\times0.8$ mm$^3$. Epoxy and copper leads were used for TTO contacts. The relative error in our measurement for thermopower was below 5\% based on the Ni standard measured under identical conditions. Sample dimensions were measured by an optical microscope Nikon SMZ-800 with resolution of 10 $\mu$m.

\section{RESULTS AND DISCUSSIONS}

\begin{table}
\caption{\label{tab}The actual chemical composition and lattice parameter $a$ for Cu$_{6-x}$Fe$_{4+x}$Sn$_{12}$Se$_{32}$ ($x$ = 0, 1, 2).}
\begin{ruledtabular}
\begin{tabular}{lll}
  Nominal & Actual & $a$ ({\AA})\\
  \hline
  Cu$_6$Fe$_4$Sn$_{12}$Se$_{32}$ & Cu$_{5.9(2)}$Fe$_{4.0(1)}$Sn$_{13(1)}$Se$_{32(1)}$ & 10.774\\
  Cu$_5$Fe$_5$Sn$_{12}$Se$_{32}$ & Cu$_{5.3(2)}$Fe$_{3.6(2)}$Sn$_{14(1)}$Se$_{31(1)}$ & 10.786\\
  Cu$_4$Fe$_6$Sn$_{12}$Se$_{32}$ & Cu$_{5.1(2)}$Fe$_{4.2(2)}$Sn$_{13(1)}$Se$_{32(1)}$ & 10.796
\end{tabular}
\end{ruledtabular}
\end{table}

\begin{figure}
\centerline{\includegraphics[scale=0.92]{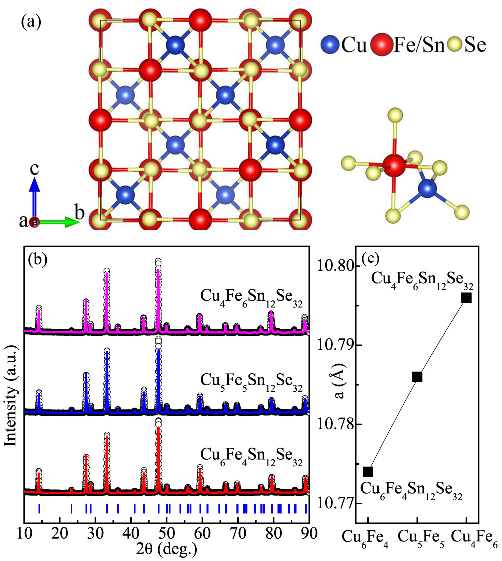}}
\caption{(Color online) (a) Crystal structure of selenospinel Cu$_{6-x}$Fe$_{4+x}$Sn$_{12}$Se$_{32}$ ($x$ = 0, 1, 2). (b) Powder x-ray
diffraction patterns and (c) the evolution of lattice parameter $a$.}
\end{figure}

Figure 1(a) shows the crystal structure of selenospinel Cu$_{6-x}$Fe$_{4+x}$Sn$_{12}$Se$_{32}$ ($x$ = 0, 1, 2). The spinel-type compounds have a general formula of AB$_2$X$_4$, where A and B are metals encapsulated by tetrahedra and octahedra, and X is oxygen or chalcogen elements. Herein, the unit cell contains numerous 54 atoms [Fig. 1(a)] with A-site Cu-deficiencies; Cu atoms are coordinated tetrahedrally by X = Se; B-site Fe/Sn atoms are randomly distributed in the same crystallographic site and are coordinated octahedrally by Se \cite{Suekuni3}. We note that Cu$_y$Fe$_4$Sn$_{12}$Se$_{32}$ is stable at $y$ = 6 \cite{Suekuni4,Suekuni5}. Figure 1(b) shows the structural refinement of powder XRD for Cu$_{6-x}$Fe$_{4+x}$Sn$_{12}$Se$_{32}$ ($x$ = 0, 1, 2), indicating that all reflections can be well indexed in the $Fd\overline{3}m$ space group. No impurity peaks are observed, confirming higher tolerance of Cu deficiencies in Cu$_{6-x}$Fe$_{4+x}$Sn$_{12}$Se$_{32}$. The determined lattice parameter $a$ increases slightly from 10.774(2) to 10.796(2) {\AA} with increasing $x$ [Fig. 1(c)]. The value of $a$ is close to the previous result for Cu$_6$Fe$_4$Sn$_{12}$Se$_{32}$ \cite{Suekuni4}, whilst the rise in $x$ suggests larger ionic size of Fe than Cu ions.

\begin{figure}
\centerline{\includegraphics[scale=1]{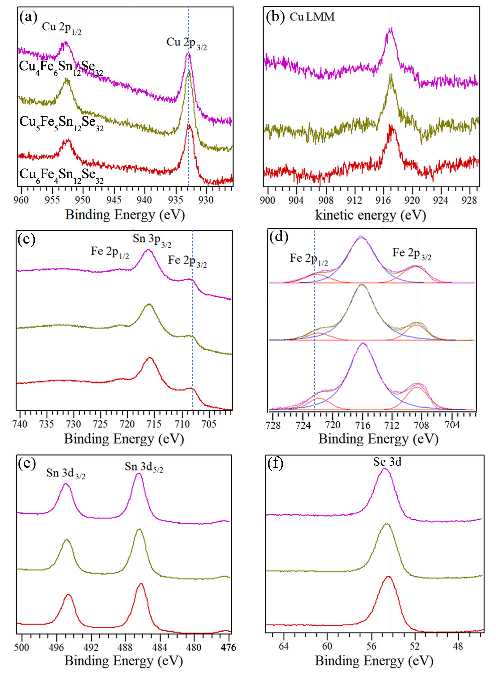}}
\caption{(Color online). Typical XPS spectra for the indicated elements of Cu$_6$Fe$_4$Sn$_{12}$Se$_{32}$ (bottom), Cu$_5$Fe$_5$Sn$_{12}$Se$_{32}$ (middle), and Cu$_4$Fe$_6$Sn$_{12}$Se$_{32}$ (top) samples.}
\label{RST}
\end{figure}

To investigate the valencies of interpolated cations, we measured the XPS where all three samples show similar valence states for each elements. As shown in Fig 2(a), the Cu $2p_{3/2}$ binding energy of 932.9 eV suggest the Cu in metallic state Cu$^0$ or Cu$^{1+}$. Auger spectra shows Cu LMM at kinetic energy of 916.8 eV [Fig. 2(b)], which exclude the possibility of Cu$^0$ and directly confirm Cu$^{1+}$ state. The highest intensity of Fe $2p$ are coincidentally overlapping with Sn $3p_{3/2}$ [Fig. 2(c)]. By deconvoluting each components, the binding energy of Fe $2p_{3/2}$ is located at 708.2 eV and shifted toward high binding energy about 1.5 eV when compare to the binding energy of pure metal Fe $2p_{3/2}$ at 706.7 eV [Fig. 2(d)]. Compared to the binding energy of Sn $3d_{5/2}$ at 485.2 eV for pure metal, the binding energy of Sn $3d_{5/2}$ is located at 486.5 eV [Fig. 2(e)] and shifted toward high binding energy about 1.3 eV. These observations suggest that the electropositive Fe [mostly Fe$^{2+}$] and Sn [mostly Sn$^{4+}$] ions in these compounds \cite{Fe,Sn}. Moreover, the binding energy of Se $3d_{5/2}$ is located at 54.5 eV and shifted toward to lower binding energy about -0.9 eV when compare to the binding energy of pure metal Se $3d_{5/2}$ at 55.4 eV [Fig. 2(f)], suggesting that the electronegative Se [mostly Se$^{2-}$] ions in the compounds. Subtle difference in binding energy of Sn and Se indicates slight local structural change induced by Fe substitution, calling for further synchrotron x-ray diffraction study on the site occupancies.

Temperature dependence of electrical resistivity $\rho(T)$ for Cu$_{6-x}$Fe$_{4+x}$Sn$_{12}$Se$_{32}$ ($x$ = 0, 1, 2) is depicted in Fig.3(a), showing an obvious semiconducting behavior. The value of room temperature resistivity ($\rho_{\textrm{300K}}$) is about
2.2$\times$$10^{-4}$ $\Omega$ m for Cu$_6$Fe$_4$Sn$_{12}$Se$_{32}$, which monotonically increases to 8.0$\times$$10^{-4}$ and
5.1$\times$$10^{-3}$ $\Omega$ m for $x$ = 1 and 2, respectively. In such a complex disordered system, the electronic conduction mechanism is of high interest. In general, three typical models are considered to describe the semiconducting behavior: (i) thermally activated model $\rho(T) = \rho_0 exp(E_\rho/k_\textrm{B}T)$, where $E_\rho$ is activation energy and $k_\textrm{B}$ is the Boltzmann constant; (ii) adiabatic small polaron hopping model $\rho(T) = AT exp(E_\rho/k_\textrm{B}T)$; and (iii) variable-range hopping (VRH) model $\rho(T) = \rho_0 exp(T_0/T)^{\nu}$ \cite{CaCoO,Austin,YL1,YL2,YL3,YL4,Rong,Fried}, where $T_0$ is a characteristic temperature and is related to density of states available at the Fermi level and carrier localization length, $\nu$ depends on the dimensionality. To understand the transport mechanism, it is necessary to fit the temperature-dependent resistivity curves based on these three formulas.

\begin{figure}
\centerline{\includegraphics[scale=1]{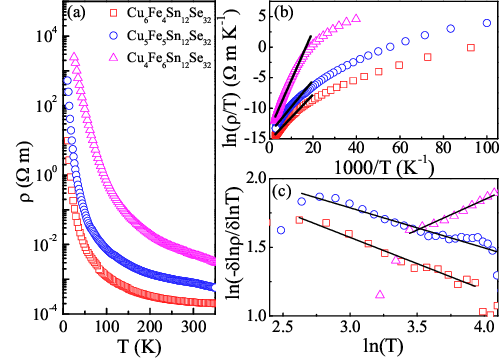}}
\caption{(Color online). (a) Temperature dependence of electrical resistivity $\rho(T)$ for the indicated samples. (b) ln($\rho/T$) versus
1000/$T$ curves fitted by the adiabatic small polaron hopping model $\rho(T) = ATexp(E_\rho/k_\textrm{B}T)$. (c) ln(-$\delta$ln$\rho$/$\delta$ln$T$) versus ln($T$), where the slopes give the exponents for variable-range hopping model $\rho(T) = \rho_0 exp(T_0/T)^{\nu}$.}
\label{RST}
\end{figure}

Figure 3(b) exhibits the fitting result of the adiabatic small polaron hopping model from 50 to 350 K. The extracted activation energy
$E_\rho$ is about 32.3(4) meV for $x$ = 0, in agreement with the previous result \cite{Suekuni4}, and gradually increases to 37.3(5) and
69.8(9) meV for $x$ = 1 and 2, respectively. However, the $\rho(T)$ curves can also be fitted by the thermally activated model (see
discussion below). With decreasing temperature, the resistivity increases abruptly below 50 K, indicating a strong localization regime with a VRH conductivity, $\rho(T) = \rho_0 exp(T_0/T)^{\nu}$. We use the logarithmic-derivative method for an accurate estimate of the $\nu$ exponents, i.e. plot ln(-$\delta$ln$\rho$/$\delta$ln$T$) versus ln($T$) [Fig. 3(c)], where the slopes of the fit give the values of exponent $\nu$. The obtained $\nu$ $(0.29 \thicksim 0.44)$ situates between 0.25 for the Mott's VRH and 0.5 for the Efros-Shklovskii-type VRH conductivity.

\begin{table}
\caption{\label{tab} Variable range hopping exponents and parameters from specific heat measurements for indicated samples.}
\begin{ruledtabular}
\begin{tabular}{lllll}
  Sample & $\nu$ & $\beta$ & $\Theta_\textrm{D}$ & $\nu_\textrm{s}$\\
   && (mJ mol$^{-1}$ K$^{-4}$) & (K) & (m s$^{-1}$) \\
  \hline
  Cu$_6$Fe$_4$Sn$_{12}$Se$_{32}$ & 0.38(2) & 87(1) & 107(1) & 1030 \\
  Cu$_5$Fe$_5$Sn$_{12}$Se$_{32}$ & 0.29(2) & 53(1) & 125(1) & 1210 \\
  Cu$_4$Fe$_6$Sn$_{12}$Se$_{32}$ & 0.44(3) & 81(1) & 109(1) & 1050
\end{tabular}
\end{ruledtabular}
\end{table}

To distinguish the thermally activated model and polaron hopping model, we further measured temperature-dependent thermopower $S(T)$. The
$S(T)$ exhibits large positive values in the whole temperature range [Fig. 4(a)], indicating dominant hole-type carriers. The room
temperature value $S_{\textrm{300K}}$ of $x$ = 0 is about 180 $\mu$V K$^{-1}$; the large value of $S$ is attributed to the small number of holes. The rise in $x$ further increases thermopower, reaching $S_{\textrm{300K}}$ = 360 $\mu$V K$^{-1}$ for $x$ = 2. The $S(1000/T)$ curves of all samples show similar shape [Fig. 4(b)] and can be fitted with the equation
$S(T) = (k_\textrm{B}/e)(\alpha+E_\textrm{S}/k_\textrm{B}T)$ from 50 to 350 K \cite{Austin}, where $E_\textrm{S}$ is activation energy and $\alpha$ is constant. The obtained activation energy for thermopower, $E_\textrm{S}$ (3.2 $\sim$ 11.5 meV), are much smaller than those for conductivity, $E_\rho$ (32.3 $\sim$ 69.8 meV), as shown in Fig. 4(c). This large discrepancy between $E_\textrm{S}$ and $E_\rho$ indicates a polaron transport mechanism of carriers. According to the polaron model, the $E_\textrm{S}$ is the energy required to activate the hopping of carriers, while $E_\rho$ is the sum of the energy needed for the creation of carriers and activating the hopping of carriers \cite{Austin}. Therefore, $E_\textrm{S}$ is smaller than $E_\rho$. The weak temperature-dependent $S(T)$ at high temperatures also supports the small polaron conduction, however, the $S(T)$ changes its slope at lower temperatures where it evolves into the VRH conduction.

\begin{figure}
\centerline{\includegraphics[scale=1]{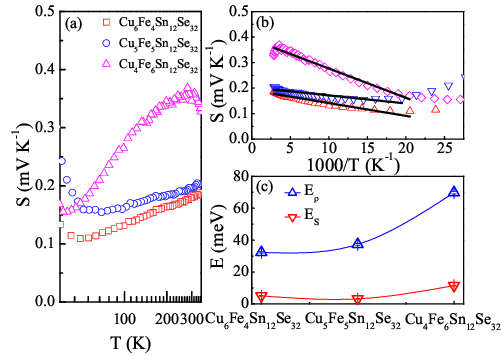}}
\caption{(Color online) (a) Temperature dependence of thermopower $S(T)$ for the indicated samples. (b) $S(T)$ vs 1000/$T$ curves fitted
using $S(T) = (k_\textrm{B}/e)(\alpha+E_\textrm{S}/k_\textrm{B}T)$ from 50 to 350 K. (c) The evolution of $E_\rho$ and $E_\textrm{S}$.}
\label{MTH}
\end{figure}

Figure 5(a) shows the temperature dependence of thermal conductivity $\kappa(T)$ for Cu$_{6-x}$Fe$_{4+x}$Sn$_{12}$Se$_{32}$ ($x$ = 0, 1,
2). In general, $\kappa_{\textrm{total}} = \kappa_\textrm{e} + \kappa_{\textrm{L}}$, consists of the electronic part $\kappa_\textrm{e}$
and the phonon term $\kappa_{\textrm{L}}$. Herein, the $\kappa_\textrm{e}$ estimated from the Wiedemann-Franz law is negligibly small due
to large electrical resistivity, indicating a predominantly phonon contribution. At room temperature, the $\kappa$ shows relatively low
value of 1.32 $\sim$ 1.57 W K$^{-1}$ m$^{-1}$, arising from its structural complexity, i.e., large Cu deficiencies and Fe/Sn site disorder
as phonon scattering centers. Interestingly, the low-temperature $\kappa(T)$ follows a quasi-linear $T$-dependence [inset in Fig. 5(a)].
This deviates from the common $\kappa$ $\sim$ $T^3$ usually observed in bulk crystals or thin films \cite{Toulokian,McConnell}, implying
nanostructural differences that are induced by different vacancies in particular grains and associated phonon frequency changes
\cite{WangZ,ZhaoH}. The Fe-substituted samples show slightly larger values of $\kappa(T)$ when compared to Cu$_6$Fe$_4$Sn$_{12}$Se$_{32}$.
This also leads to the increase in phonon mean free path $l_\kappa$ [Fig. 5(b)] estimated from the heat capacity and thermal conductivity via $\kappa_\textrm{L} = C_\textrm{p}\nu_\textrm{s}l_\kappa/3$, where $\kappa_\textrm{L}$ is the lattice thermal conductivity, $C_\textrm{p}$ is the heat capacity, $\nu_\textrm{s}$ is the average sound speed calculated via Debye temperature \cite{SV,QH}, and $l_\kappa$ is the phonon mean free path.

\begin{figure}
\centerline{\includegraphics[scale=1]{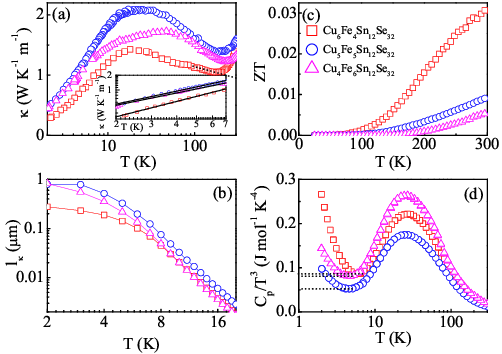}}
\caption{(Color online) Temperature dependence of (a) thermal conductivity $\kappa(T)$, (b) phonon mean free path $l_\kappa$, (c) the figure of merit $ZT$, and (d) specific heat $C_\textrm{p}$ divided by $T^3$ for the indicated samples.}
\label{MTH}
\end{figure}

The figure of merit $ZT$ shows maxima around the room temperature, reaching $\sim$ 0.03 for Cu$_6$Fe$_4$Sn$_{12}$Se$_{32}$ at 300 K
[Fig. 5(c)]. Note that the increase of $\kappa$ above 200 K can be due to radiation and as a consequence the real $ZT$ can be $\sim$ 0.044 for Cu$_6$Fe$_4$Sn$_{12}$Se$_{32}$ at 300 K. Although the thermopower is enhanced by Fe-substitution, the electrical resistivity and thermal conductivity are also increased in Fe-substituted samples. Further efforts in nanostructuring of single crystal alloys and/or doping with heavier element would be helpful to obtain better thermoelectricity performance at room temperature.

Temperature dependence of specific heat $C_\textrm{p}$ divided by $T^3$ of the indicated samples is depicted in Fig. 5(d). Two interesting features are observed: a hump around 25(5) K and, it diverges from the Debye $T^3$ law at low temperatures, similar to those observed in amorphous solids arising from the enhanced density of states of acoustic phonons caused by the disordered structure \cite{1,2,3,4,5,6}. The coefficients $\beta$ = 53(1) $\sim$ 87(1) mJ mol$^{-1}$ K$^{-4}$ were estimated from the local minimum values of $C_\textrm{p}/T^3$, as shown by dashed lines in Fig. 4(d). The derived Debye temperature $\Theta_\textrm{D}$ = 107(1) $\sim$ 125(1) K by using the equation $\Theta_\textrm{D} = [12\pi^4NR/(5\beta)]^\frac{1}{3}$, which implies the average sound velocity of $\nu_\textrm{s} \approx 1030 \sim 1210$ m s$^{-1}$ [Table II]. We should note that the large values of $\gamma$ can also be from magnetic spin glassy state in this high Fe-content disordered system. Further synchrotron local structural (pair distribution function analysis) and neutron scattering experiments will be helpful to unveil its origin.

\section{CONCLUSIONS}

In summary, we synthesized and studied single crystals of Cu$_{6-x}$Fe$_{4+x}$Sn$_{12}$Se$_{32}$ ($x$ = 0, 1, 2). Electronic transport
mechanism on cooling is dominated by polaron effect down to 50 K. On further cooling the VRH mechanism dominates. The Fe substitution enhances the thermopower possibly due to an increase of the slope at and shift of the density of states near the Fermi level. The figure of merit $ZT$ reaches $\sim$ 0.03 - 0.044 around the room temperature for Cu$_6$Fe$_4$Sn$_{12}$Se$_{32}$, calling for further carrier optimization.

\section*{Acknowledgements}

Work at BNL is supported by the Office of Basic Energy Sciences, Materials Sciences and Engineering Division, U.S. Department of Energy
(DOE) under Contract No. DE-SC0012704. A portion of this work was performed at the National High Magnetic Field Laboratory, which was
supported by the National Science Foundation Cooperative Agreement No. DMR-1644779 and the State of Florida.

$^{*}$Present address: Los Alamos National Laboratory, Los Alamos, NM 87545, USA.\\

\end{document}